\documentclass[preprintnumbers,amsmath,two column]{revtex4}
\usepackage{graphicx}
\usepackage{dcolumn}
\usepackage{bm}

\begin{document}

\title{Ground-state properties of the two-site Hubbard-Holstein model: an exact
solution}
\author{Yu-Yu Zhang$^{1}$, Tao Liu$^{2}$, Qing-Hu Chen$^{1,3,\dag}$, Xiaoguang Wang$%
^{1}$, Kelin Wang$^{2,4}$}

\address{$^{1}$Department of Physics, Zhejiang University,
 Hangzhou 310027, P. R. China.\\
$^{2}$Department of Physics, Southwest University of Science and
Technology, Mianyang 621010, P. R. China.\\
$^{3}$Center for Statistical and Theoretical Condensed Matter
Physics, Zhejiang Normal University, Jinhua 321004, P. R. China \\
$^{4}$Department of Mordern Physics, University of Science and
Technology of China, Hefei 230026, P. R. China.}

\date{\today}

\begin{abstract}
We revisit the two-site Hubbard-Holstein model by using extended
phonon coherent states. The nontrivial singlet bipolaron is studied
exactly in the whole coupling regime. The  ground-state (GS) energy
and the double occupancy probability are  calculated. The linear
entropy is exploited successfully to quantify bipartite entanglement
between electrons and their environment phonons, displaying a
maximum entanglement of the singlet-bipolaron in strong coupling
regime. A dramatic drop in the crossover regime is observed in the
GS fidelity and its susceptibility. The bipolaron properties is also
characterized classically by correlation functions. It is found that
the crossover from a two-site to single-site bipolaron is more
abrupt and shifts to a larger electron-phonon coupling strength as
electron-electron Coulomb repulsion increases.
\end{abstract}

\maketitle

\section{introduction}

It has been demonstrated for more than one decade that the polaronic
effects observed in High-$T_{c}$ superconducting
cuprates~\cite{ASAlexandrov} and the colossal magnorestive
manganites~\cite{AJMilis} are relevant to the electron-phonon (e-ph)
coupling in these systems. The famous Holstein molecular crystal
model~\cite{THolstein}, where electrons are coupled with local
phonons, has then revived in recent years. More recently, to include
the electron-electron (e-e) Coulomb repulsion interactions, the
Hubbard-Holstein (HH) model~\cite{Fehske} for strongly correlated
electron systems has made significant progress in understanding
many-body aspects. As is well known, an exact solution to this model
in the thermodynamic limit is impossible, and many approximate
approaches are then employed. The two-site HH model, where two
electrons hop between two adjacent lattice sites, can be solved
analytically\cite{Berciu}. It is not only a prototype of the HH
model but also helpful for the better understanding to the polaron
and bipolaron behavior in infinite lattices.

Recently, an contemporary alternative characterization of the ground
state (GS) properties has been focused on quantum information tools
in terms of quantum
entanglement~\cite{Scott,William,Vidal,Gu,chenqh} and
fidelity~\cite{Quan,Zanardi,Vezzani,Min,Shu,Buon,Chen}. These
studies will establish somewhat interesting understanding from the
field of quantum information theory to condensed matter
physics~\cite{Dunning,Norman}. Usually, it is hard to calculate
these quantities due to the lack of knowledge on the exact GS wave
function.

The two-site HH model has been previously addressed by means of different
methods, such as variational method \cite{ANDas,Acquarone}, perturbation
theory~\cite{JC}, and numerical diagonalization\cite{Ranninger,EVLde}. In
spite of these efforts there remains poor convergence in the intermediate
coupling regime. A reliable treatment in the whole coupling regime is still
needed. Recently, Berciu derived all Greens's function analytically for the
two-site HH model in terms continued fractions\cite{Berciu}.

In this work, by using extended bosonic coherent states
~\cite{chen,Han,liu}, we develop a new exact technique to deal with
the two-site HH model. The wave function is proposed explicitly, by
which many quantities can be calculated directly. The paper is
organized as follows. In Sec II we introduce the model and describe
the approach. In Sec III we calculate the linear entropy , the GS
fidelity and its susceptibility to study crossover properties from
quantum information perspective. The static correlation function is
also evaluated to analyze the GS properties. The main conclusions
are given in the last section.

\section{ Model Hamiltonian and exact solution}

The Hamiltonian of the two-site HH model takes the form
\begin{eqnarray}  \label{hamiltonian}
H&=&\sum_{i,\sigma}\varepsilon n_{i \sigma}
-\sum_{\sigma}t(c_{1\sigma}^{\dagger}c_{2\sigma}+c_{2\sigma}^{\dagger}c_{1%
\sigma})  \nonumber \\
&+&U\sum_{i}n_{i\uparrow}n_{i\downarrow}+Vn_{1}n_{2}
+g_{1}\omega_{0}\sum_{i,\sigma}n_{i,\sigma}(b_{i}+b_{i}^{\dagger})  \nonumber
\\
&+&g_{2}\omega_{0}\sum_{i,\sigma}n_{i,\sigma}
(b_{i+\delta}+b_{i+\delta}^{\dagger})+\omega_{0}\sum_{i}b_{i}^{\dagger}b_{i},
\end{eqnarray}
where $i(=1$ or $2$) denotes the label of sites. $i+\delta=2$ for $i=1$ and
vice versa. $c_{i\sigma}(c_{i\sigma}^{\dagger})$ is the annihilation
(creation) operator for the electrons and $n_{i,\sigma}(=c_{i\sigma}^{%
\dagger}c_{i\sigma})$ is the corresponding number operator at site $i$ with
spin $\sigma$, besides $n_{i}=n_{i\uparrow}+n_{i\downarrow}$. $\varepsilon$
is the unperturbed site potential. $t$ is the usual hopping integral. $U$
and $V$ denote the on-site and inter-site Coulomb repulsion between
electrons respectively. $g_{1}$ and $g_{2}$ denote the on-site and
inter-site e-ph coupling parameters. $b_{i}(b_{i}^{\dagger})$ is the
annihilation (creation) operator for phonons corresponding to interatomic
vibrations at site $i$. $\omega_{0}$ is the phonon frequency and is set unit
for convenience in the following.

Introducing new phonon operators $a=(b_{1}+b_{2})/\sqrt{2}$ and $%
d=(b_{1}-b_{2})/\sqrt{2}$, the Hamiltonian ~(\ref{hamiltonian}) can be
written into two independent parts ($H=H_{d}+H_{a}$),
\begin{eqnarray}  \label{hamiltonian d}
H_{d}&=&d^{\dagger}d+\sum_{i,\sigma}\varepsilon n_{i
\sigma}-\sum_{\sigma}t(c_{1\sigma}^{\dagger}c_{2\sigma}+c_{2\sigma}^{%
\dagger}c_{1\sigma})  \nonumber \\
&+&U\sum_{i}n_{i\uparrow}n_{i\downarrow}+Vn_{1}n_{2}  \nonumber \\
&+&g_{-}(n_{1}-n_{2})(d+d^{\dagger})-n^{2}g_{+}^{2}
\end{eqnarray}
and
\begin{equation}
H_{a} = \tilde{a}^{\dagger}\tilde{a},
\end{equation}
where $\tilde{a}^{\dagger}=a^{\dagger}+ng_{+}$, $\tilde{a}=a+ng_{+}$, $%
g_{+}=(g_{1}+g_{2})/\sqrt{2}$, and $g_{-}=(g_{1}-g_{2})/\sqrt{2}$. $H_{a}$
describes a shifted oscillator and represents lowering of energy, which is a
constant motion. $H_{d}$ represents an effective e-ph coupling system which
phonons are coupled linearly with the electrons.

In the effective Hamiltonian $H_{d}$, among four different
bipolaronic states, such as singlet bipolaronic states, singlet
Anderson bipolaronic states, singlet bipolaronic states
(antibonding), and triplet states, the singlet bipolaronic state is
the most nontrivial to any approaches. We will focus on this state
in this paper.

For three singlet normalized electronic states $c_{1\uparrow}^{\dagger}c_{1%
\downarrow}^{\dagger}|0\rangle_{e}$, $c_{2\uparrow}^{\dagger}c_{2%
\downarrow}^{\dagger}|0\rangle_{e}$ and $\frac{1}{\sqrt{2}}%
(c_{1\uparrow}^{\dagger}c_{2\downarrow}^{\dagger}-c_{1\downarrow}^{%
\dagger}c_{2\uparrow}^{\dagger})|0\rangle_{e}$, the singlet bipolaronic
state wave function $|\psi\rangle$ can be expressed as
\begin{eqnarray}  \label{SB}
|\psi\rangle&=& |\varphi_{1}\rangle
c_{1\uparrow}^{\dagger}c_{1\downarrow}^{\dagger}|0\rangle_{e}+
|\varphi_{2}\rangle c_{2\uparrow}^{\dagger}c_{2\downarrow}^{\dagger}
|0\rangle_{e}  \nonumber \\
&+&|\varphi_{3}\rangle\frac{1}{\sqrt{2}}(c_{1\uparrow}^{\dagger}c_{2%
\downarrow}^{\dagger}-c_{1\downarrow}^{\dagger}c_{2\uparrow}^{\dagger})|0%
\rangle_{e}
\end{eqnarray}
where $|\varphi_{1}\rangle$, $|\varphi_{2}\rangle$ and $|\varphi_{3}\rangle$
correspond to phonon states. Inserting it into a schr\"{o}dinger equation
for the effective Hamiltonian in Eq. (\ref{hamiltonian d}), three equations
are derived by comparing the coefficients of $c_{1\uparrow}^{\dagger}c_{1%
\downarrow}^{\dagger}|0\rangle_{e}$, $c_{2\uparrow}^{\dagger}c_{2%
\downarrow}^{\dagger}|0\rangle_{e}$ and $\frac{1}{\sqrt{2}}%
(c_{1\uparrow}^{\dagger}c_{2\downarrow}^{\dagger}-c_{1\downarrow}^{%
\dagger}c_{2\uparrow}^{\dagger})|0\rangle_{e}$,
\begin{equation}  \label{eq2}
[A^{\dagger}A+2\varepsilon-4(g_{1}^{2}+g_{2}^{2})+U]|\varphi_{1}\rangle-%
\sqrt{2}t|\varphi_{3}\rangle=E|\varphi_{1}\rangle
\end{equation}
\begin{equation}  \label{eq3}
[B^{\dagger}B+2\varepsilon-4(g_{1}^{2}+g_{2}^{2})+U]|\varphi_{2}\rangle-%
\sqrt{2}t|\varphi_{3}\rangle=E|\varphi_{2}\rangle
\end{equation}
\begin{equation}  \label{eq4}
[d^{\dagger}d+2\varepsilon-2(g_{1}+g_{2})^{2}+V]|\varphi_{3}\rangle-\sqrt{2}%
t(|\varphi_{1}\rangle+|\varphi_{2}\rangle)=E|\varphi_{3}\rangle
\end{equation}
where we have used two displacement transformation $A^{\dagger}=d^{%
\dagger}+2g_{-}$ and $B^{\dagger}=d^{\dagger}-2g_{-}$. Note that the linear
term for the phonon operator $d(d^{+})$ is removed, and two new free bosonic
field with operator $A(A^{+})$ and $B(B^{+})$ appear. In the next step, we
naturally choose the basis in terms of these new operator, instead of $%
d(d^{+})$, by which the phonon states $|\varphi_{1}\rangle$ and $%
|\varphi_{2}\rangle$ can be expanded in complete basis
${|n\rangle_{A}}$ and ${|n\rangle_{B}}$ respectively, where
$|n\rangle_{A}$ and $|n\rangle_{B}$ are Fock states of the new
bosonic operators
\begin{eqnarray}  \label{function1}
&|\varphi_{1}\rangle&=\sum_{n=0}^{N_{tr}}c_{n}|n\rangle_{A}  \nonumber \\
&=&\sum_{n=0}^{N_{tr}}c_{n}\frac{1}{\sqrt{n!}}(d^{%
\dagger}+2g_{-})^{n}e^{-2g_{-}d^{\dagger}-2g_{-}^{2}}|0\rangle_{ph}
\end{eqnarray}

\begin{eqnarray}  \label{function2}
&|\varphi_{2}\rangle&=\sum_{n=0}^{N_{tr}}d_{n}|n\rangle_{B}  \nonumber \\
&=&\sum_{n=0}^{N_{tr}}d_{n}\frac{1}{\sqrt{n!}}(d^{%
\dagger}-2g_{-})^{n}e^{2g_{-}d^{\dagger}-2g_{-}^{2}}|0\rangle_{ph}
\end{eqnarray}

As we know that the vacuum state $\left| 0\right\rangle _{A(B)}$ is
just a bosonic coherent-state in $d(d^{+})$ with an eigenvalue
$2g(-2g_{-})$\cite {chen,Han,liu}. So this new basis is
overcomplete, and actually does not involve any truncation in the
Fock space of $d(d^{+})$, which highlights the present approach. It
is also clear that many-body correlations for bosons
are essentially included in extended coherent states (\ref{function1}) and (%
\ref{function2}). As usual, the phonon state $|\varphi_{3}\rangle$ is
expanded in a complete basis ${|n\rangle}$, which is the Fock state of $%
d(d^{+})$
\begin{equation}  \label{function3}
|\varphi_{3}\rangle=\sum_{n=0}^{N_{tr}}f_{n}|n\rangle
\end{equation}

Substituting these phonon states~(\ref{function1}),~(\ref{function2}), and~(%
\ref{function3}) into Eqs.~(\ref{eq2}),(~\ref{eq3}) and ~(\ref{eq4})
and left multiplying state $_{A}\langle m|$, $_{B}\langle m|$,
$\langle m|$, respectively, the equations can be written as
$3N_{tr}\times3N_{tr}$ matrix, and the $m^{\prime}$th row is written
as
\begin{equation}  \label{eq5}
[m+2\varepsilon-4(g_{1}^{2}+g_{2}^{2})+U]c_{m}-\sqrt{2}t\sum_{n}{_{A}\langle
m}|n\rangle f_{n}=Ec_{m}
\end{equation}
\begin{equation}  \label{eq6}
[m+2\varepsilon-4(g_{1}^{2}+g_{2}^{2})+U]d_{m}-\sqrt{2}t\sum_{n}{_{B}\langle
m}|n\rangle f_{n}=Ed_{m}
\end{equation}

\begin{eqnarray}  \label{eq7}
&[m&+2\varepsilon-2(g_{1}+g_{2})^{2}+V]f_{m} -\sqrt{2}t(\sum_{n}c_{n}\langle
m|n\rangle_{A}  \nonumber \\
&+&\sum_{n}d_{n}\langle m|n\rangle_{B})=Ef_{m}
\end{eqnarray}

where $ _{A}\langle m|n\rangle =\langle
m|n\rangle_{B}=(-1)^{n}D_{mn}$, and $_{B}\langle m|n\rangle
=\langle m|n\rangle_{A}=(-1)^{m}D_{mn} $, with $
D_{mn}=e^{-2g_{-}^{2}}\sum_{i=0}^{{min}[m,n]}(-1)^{-i}\frac{\sqrt{
m!n!}(2g_{-})^{m+n-2i}}{(m-i)!(n-i)!i!}$.

The eigenvalues $E$ and eigenvectors with coefficients ${c_{n}}$,
${d_{n}}$
and $f_{n}$ can be exactly solved by diagonalizing the above $%
3N_{tr}\times3N_{tr}$ matrix numerically.

To obtain the true exact results, in principle, the truncated number
$N_{tr}$ should be taken to infinity. Fortunately, finite terms of
the singlet state in Eq.(4) are sufficient to give very accurate
results in the whole parameter range. It should be noted that in the
exact diagonalization in Fock space of original phonon state $d$
\cite{EVLde}, a considerable large phonon number is needed to give
reasonably good results. We believe that we have exactly solved this
model numerically.

\begin{figure}[tbp]
\includegraphics[scale=0.68]{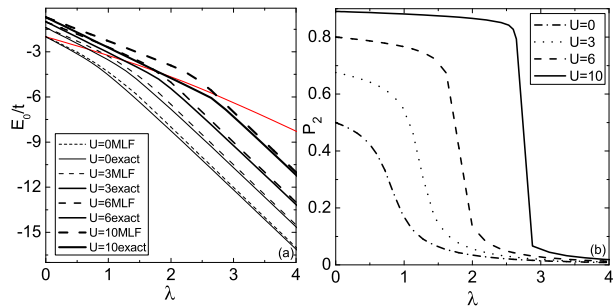}
\caption{(a) GS energy $E_{0}/t$ of exact solution (solid line)
and MLFS transformation (dashed line) vs $\lambda$ for $U=0,3,6$
and $10$. The red line represents the twice polaron GS energy,
which then separates the polaron and
bipolaron regime. (b) The probability of the two-site bipolarons $P_2$ for $%
U=0$, $3$, $6$, and $10$. The other parameters are chosen: $V=0$, $t=2.0$, $%
\varepsilon=0$.}
\label{SB}
\end{figure}

To show the effectiveness of the present approach, we first calculate the GS
energy. Fig. 1(a) presents the GS energy $E_{0}/t$ as a function of the
on-site effective coupling strength $\lambda=g_{1}^{2}/t$ by setting $%
g_{2}=0 $ conveniently. The results for the energy by the variational method
based on the modified Lang-Firsov transformation with a squeezing phonon
state transformations (MLFS)~\cite{ANDas} are also list. It is observed that
the present results are lower than the MLFS results \cite{ANDas}, especially
in the intermediate coupling regime. Comparing with the Fig. 1(a) in Ref.
\cite{Berciu}, we find that the present results for the GS energy are
consistent with those from the lowest pole of a Green's function.

\section{ground state properties}

\subsection{Crossover from two-site to single-site bipolarons}

As shown in Fig. 1(a) that there is two regimes distinctly. In the Ref. \cite
{Berciu}, the energy vs $\lambda$ curves in two regimes are fitted by two
functions, and the abrupt drop of the first-order derivative signals the
crossover regime from two-site bipolarons to single-site bipolarons. We will
propose a quantitative criterion. Because the exact wave function in the
present technique is explicitly given, we can calculate the probability of
the system that two electrons are in two sites by Eq. (4) directly.
\begin{equation}
P_2=Re\langle\varphi_{3}|\varphi_{3}\rangle
\end{equation}
The probability of the two-site bipolarons $P_2$ is shown in Fig.
1(b). At the weak coupling, the two electrons prefer to stay in two
sites. As the coupling strength increases, the two electrons tend to
occupy the single site. Strictly speaking, at weak (strong)
coupling, it is not pure two-site (single-site) bipolarons, only
dominated two-site (single-site) bipolarons. The crossover regime is
wide for weak on-site Coulomb repulsion $U$. As $U$ increases, the
crossover shifts to larger $\lambda$ and becomes more sharp.
However, the jump of the $P_2$ is not observed, indicating unlikely
a phase transition.

\begin{figure}[tbp]
\includegraphics[width=7.8cm]{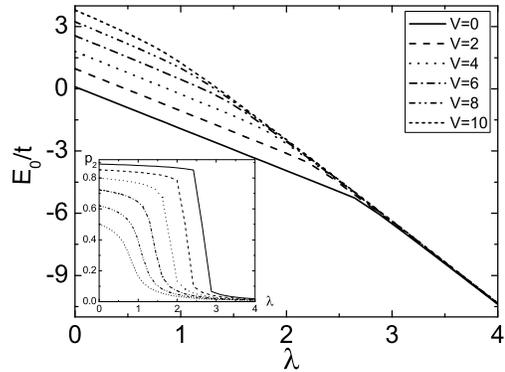}
\caption{GS energy $E_{0}/t$ and the probability of the two-site bipolarons $%
P_2$ in the inset vs $\lambda$ for $V=0,2,4,6,8,10$ by choosing $V=0.2$, $%
t=2.0$, $\varepsilon=0.8$.}
\label{p2}
\end{figure}
The inter-site Coulomb repulsion $V$ may favor the formation of dominated
single-site bipoalrons. To show this effect, we calculate the GS energy and
the probability of the two-site bipolarons for several values of $V$ for
fixed $U$, which are displayed in Fig. 2. It is clear that the crossover
shifts to smaller $\lambda$ with increasing $V$, and becomes more smooth.

\begin{figure}[tbp]
\includegraphics[scale=0.7]{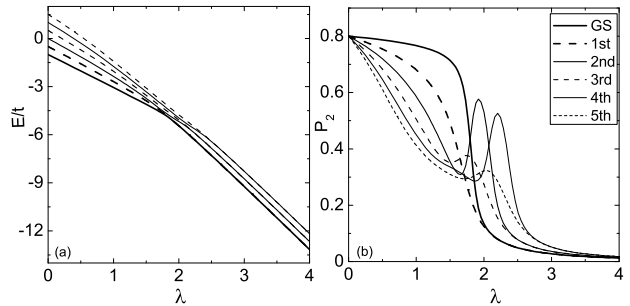}
\caption{GS energy and the former five excited energies $E/t$ (a) and the
corresponding probability $P_2$ (b) against $\lambda$ for $U=6$ by choosing $%
V=0.2$, $t=2.0$, $\varepsilon=0.8$.}
\label{excite}
\end{figure}

In our approach, we can also get the excited states and corresponding
energies. Both the GS and a few low excited state bipolaron energies for $%
U=6 $ are presented in Fig.~\ref{excite}(a). Interesting, we find
that the curves for the GS and the 1st excited state energies, the
2nd and 3rd excited state energies, 4th and 5th excited state
energies almost merge into a single line in the strong coupling
regime. The corresponding two eigenstates are almost degenerated.
Actually they are different. The difference is too small to be
observed. The merged point shift to larger $\lambda$ for higher
excited states. To show the bipolaron behavior in these states, we
also calculate the probability of the two-site bipolarons $P_2$ as a
function of $\lambda$, which are shown in Fig.\ref{excite}(b). The
$P_2$ vs $\lambda$ curves also merge into a single line exactly at
the same point as the energy vs $\lambda$ curves. In the GS and 1st
excited state, the value of $P_2$ decreases monotonously as
$\lambda$. However, in the other high excited states, a
non-monotonous behavior of the $P_2$ is observed. In a fixed value of $%
\lambda$, the probability of the two-site bipolarons $P_2$ determines the
positive repulsion energy for the double occupancy, the energy gained for
the two electrons in the same site, and the energy gained for the hopping of
electrons from one site to other site. To increase the energy in higher
excited states, the competition of these energies contributes the
complicated behavior shown in Fig.\ref{excite}(b).

\subsection{Linear entropy}

To investigate the crossover from two-site to single-site
bipolarons in quantum information science, we attempt to study
quantum entanglement between electrons and their surrounding
phonons by means of the linear entropy $E_{l}$~\cite{Yang,Wang},
which is an alternative measurement to indicate the entanglement
of a two-site and single-site bipolaron. It is defined as
\begin{equation}
E_{l}=1-{Tr}\rho^{2},
\end{equation}
where $\rho=Tr_{ph}(|\psi\rangle \langle\psi|)$ is the reduced density
matrix of electrons by taking partial trace over the phonon degrees of the
freedom.

The normalized GS wave function of the singlet bipolaronic state in Eq.(4)
is described as
\begin{equation}
|\psi\rangle=|\varphi_{1}\rangle|1\rangle+|\varphi_{2}\rangle|2\rangle+|%
\varphi_{3}\rangle|3\rangle,
\end{equation}
where the phonon states satisfy $\langle\varphi_{1}|\varphi_{1}\rangle+%
\langle\varphi_{2}|\varphi_{2}\rangle+\langle\varphi_{3}|\varphi_{3}%
\rangle=1 $. And the quantum states $|1\rangle$, $|2\rangle$ and $|3\rangle$
represent the normalized and orthogonalized singlet electronic states $%
c_{1\uparrow}^{\dagger}c_{1\downarrow}^{\dagger}|0\rangle_{e}$ $%
c_{2\uparrow}^{\dagger}c_{2\downarrow}^{\dagger}|0\rangle_{e}$ and $\frac{1}{%
\sqrt{2}}(c_{1\uparrow}^{\dagger}c_{2\downarrow}^{\dagger}-c_{1\downarrow}^{%
\dagger}c_{2\uparrow}^{\dagger})|0\rangle_{e}$ respectively.

The reduced density matrix $\rho$ can be derived by taking partial trace
over the phonon degrees of the freedom $\rho=\sum^{3}_{i,j=1}{%
\langle\varphi_{j}}|\varphi_{i}\rangle|i\rangle\langle j|$. Therefore the
linear entropy can be derived simply as
\begin{eqnarray}
E_{l}&=&1-[\langle\varphi_{1}|\varphi_{1}\rangle^{2}+\langle\varphi_{2}|%
\varphi_{2}\rangle^{2}+\langle\varphi_{3}|\varphi_{3}\rangle^{2}  \nonumber
\\
&+&2Re(\langle\varphi_{1}|\varphi_{2}\rangle^{2}+\langle\varphi_{1}|%
\varphi_{3}\rangle^{2}+\langle\varphi_{2}|\varphi_{3}\rangle^{2})]
\end{eqnarray}

\begin{figure}[tbp]
\includegraphics[width=6cm]{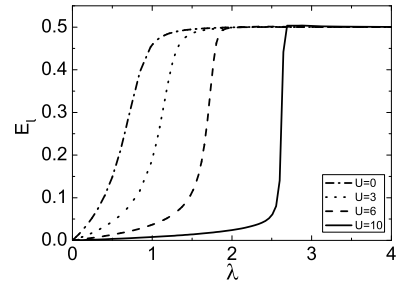}
\caption{Linear entropy $E_{l}$ vs $\lambda$ for $U=0,3,6,10$ by choosing $%
V=0.2$, $t=2.0$, $\varepsilon=0.8$. The maximum $E_l =0.5$
corresponds to the single-site bipolaron in strong coupling
region.} \label{entangl1}
\end{figure}

As plotted in Fig.\ref{entangl1}, the linear entropy $E_{l}$
increases smoothly with the e-ph coupling parameters $\lambda$ for
different e-e Coulomb repulsions $U$. In general, in weak e-ph
coupling region, two electrons tend to occupy two sites, and form
so-called two-site bipolaron, resulting a low degree of the quantum
entanglement in the original coupling strength. For a fixed $U$, as
$\lambda$ increases two electrons become to tightly interact with
the same lattice. So the entire charge of the electrons and entire
deformations of the lattices are restricted on one site, leading to
the formation of a single-site bipolaron. However, the two-site
bipolaron is known as that each electron just interacts with phonons
of its own site and then the linear entropy is expected to be much
smaller than that of the singlet-site bipolaron. As shown in
Fig.\ref{entangl1} the linear entropy reaches its maximum
$E_{l}=0.5$ in the strong coupling region, which displays that the
single-site bipolarons are maximally entangled. Obviously the
crossover point shifts to a large value of the coupling strength
$\lambda$ as $U$ increases. The similar crossover point has been
demonstrated by the probability of the two-site bipolarons presented
in Fig. 1(b). Therefore we can say that in the presence of quantum
correlations the two-site and singlet-site bipolarons are
effectively characterized to a high (low) degree of bipartite
quantum entanglement.

\subsection{ ground state fidelity}

More recently, another temporary effective quantum information
tools, i.e., the fidelity has been put forward to analyze
complicated interacting systems from the perspective of the GS wave
functions~\cite{Shu,Chen}, first identifying polaronic crossover
behaviors in bosonic system~\cite{Norman}. A simple expression of
the fidelity $F(g_{1},g_{1}+\delta g_{1})$ is given just by the
modulus of the overlap of two ground states corresponding to two
different coupling parameters $g_{1}$ and $g_{1}+\delta g_{1}$,
where $g_{1}$ is a tiny perturbation parameter~\cite{Zanardi}. It is
expected to signal a peak at the critical point. The general GS
fidelity is written as

\begin{eqnarray}
F&=&\langle\psi(g_{1})|\psi(g_{1}+\delta g_{1})\rangle  \nonumber\\
 &=&|\langle\varphi_{1}(g_{1})|\varphi_{1}(g_{1}+\delta
g_{1})\rangle+ \langle\varphi_{2}(g_{1})|\varphi_{2}(g_{1}+\delta
g_{1})\rangle  \nonumber\\
&+&\langle\varphi_{3}(g_{1})|\varphi_{3}(g_{1}+\delta
g_{1})\rangle|
\end{eqnarray}
where $|\psi(g_{1})\rangle$ and $|\psi(g_{1}+\delta g_{1})\rangle$
are two normalized GS corresponding to neighboring Hamiltonian
parameters. In our calculation $\delta g_{1}=10^{-4}$ is used. While
the fidelity susceptibility $S(g_{1})$ is regarded as a more
effective tool to detect the singularity at the crossover regime,
which reads~\cite{Cozzini}
\begin{equation}
S(g_{1})=\lim_{\delta g_{1}\longrightarrow
0}[1-F(g_{1},g_{1}+\delta g_{1})]/(\delta g_{1}^{2})
\end{equation}

\begin{figure}[tbp]
\includegraphics[width=9cm]{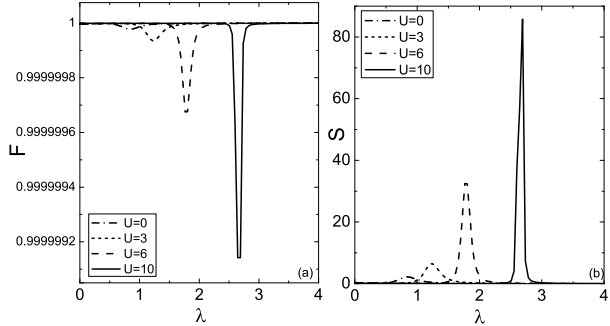}
\caption{GS fidelity $F$ and its susceptibility $S$ versus $\lambda$ for $%
U=0,3,6,10$ by choosing $V=0.2$, $t=2.0$, $\varepsilon=0.8$. A sharper drop
is shown at the crossover for larger $U=10$(solid line).}
\label{fidelity}
\end{figure}

The GS fidelity $F$ and its susceptibility $S$ as a functions of
effective coupling strength $\lambda$ ($=g_{1}^{2}/t$)
for different e-e Coulomb repulsion $U$ are displayed in Fig. \ref{fidelity}%
(a) and (b). A narrow dramatic drop is observed in the vicinity of
the bipolaron transition point as a consequence of the dramatic
change of the structure of the GS wave function, implying that the
crossover from the two-site to singlet-site bipolaron occurs. As
shown in Fig. \ref{fidelity}%
(a), the values of the GS fidelity are approximated to $1$. As is
known, for the quantum phase transition system the GS wave
functions from different sides of the level-crossing point are
almost orthogonal~\cite{Shu}. However, there is absent of the
level-crossing and then values of peaks of the GS fidelity are
around to $1$ rather than $0$ at the crossover point. Further
evidence for this crossover is given by the susceptibility $S$,
which shows a cusp
structure in the intermediate coupling regime in Fig. \ref{fidelity}%
(b). Obviously, the crossover becomes more abrupt and the critical
value of the e-ph coupling $\lambda$ where the peak appears becomes
larger as the Coulomb repulsion $U$ increases from $0$ to $10$. So
it is illustrated that the transition behavior from two-site to
singlet-site bipolaron can be effectively detected by a singularity
of the GS fidelity and its susceptibility.

\subsection{static correlation function}

Since the bipolaron crossover transition has been indeed known
from the quantum information perspective. Naturally, we seek
classically to discuss this crossover
by the means of the static on-site and inter-site correlation functions $%
\langle n_1u_1\rangle $ and $\langle n_1u_2\rangle $, which reveal
the spatial extent of lattice deformations induced by electrons
respectively. The $u_i$ denotes the lattice deformations on site
$i$ produced by the electrons and $n_i$ is the number operator of
the electrons. The GS correlation functions are written as
\begin{equation}
\langle n_1u_{1,2}\rangle =\langle n_1[\pm (d+d^{\dagger
})/2-2gn_{1,2}]\rangle   \label{corre}
\end{equation}
The positive (negative) sign and $n_1$ ($n_2$) are associated with $\langle
n_1u_1\rangle $ and $\langle n_1u_2\rangle $ respectively. By using the
exact GS wave function of the singlet bipolaronic state obtained above, the
correlation functions Eq.(\ref{corre}) can be expressed as follows
\begin{eqnarray}
&&\langle n_1u_2\rangle= -2g_{1}\sum_{m=0}^Nf_m^2-  \nonumber \\
&&\sum_{m=1}^N\sqrt{m}(c_m^{*}c_{m-1}+0.5f_m^{*}f_{m-1})  \nonumber \\
&&-\sum_{m=0}^{N-1}\sqrt{m+1}(c_m^{*}c_{m+1}+0.5f_m^{*}f_{m+1})
\end{eqnarray}
\begin{eqnarray}
\langle n_1u_1\rangle  &=&-2g_{1}\sum_{m=0}^N(4c_m^2+f_m^2)  \nonumber \\
&&+\sum_{m=1}^N\sqrt{m}(c_m^{*}c_{m-1}+0.5f_m^{*}f_{m-1})  \nonumber \\
&&+\sum_{m=0}^{N-1}\sqrt{m+1}(c_m^{*}c_{m+1}+0.5f_m^{*}f_{m+1}).
\end{eqnarray}

\begin{figure}[tbp]
\includegraphics[width=8.3cm]{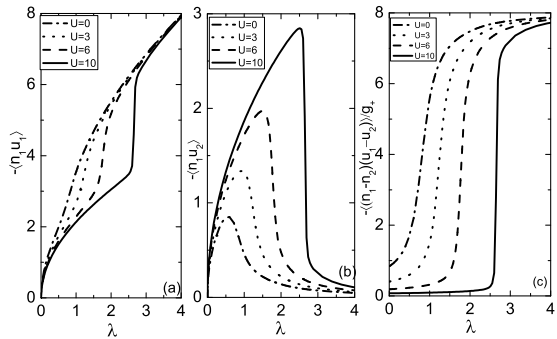}
\caption{Correlation function versus $\lambda$ (a)$-\langle
n_{1}u_{1}\rangle $, (b)$-\langle n_{1}u_{2}\rangle$, (c)$%
-\langle(n_{1}-n_{2})(u_{1}-u_{2})\rangle/g_{1}$ for $U=0,3,6,10$
by choosing $V=0.2$, $t=2.0$, $\varepsilon=0.8$. A more abrupt
crossover occurs for larger $U=10$ (solid line).}
\label{correlation}
\end{figure}

The functions $-\langle n_{1}u_{1}\rangle$ and $-\langle
n_{1}u_{2}\rangle$ against the e-ph coupling strength $\lambda$
are plotted in Figs.~\ref {correlation} (a) and (b). One can
observe that $-\langle n_{1}u_{1}\rangle$ and $-\langle
n_{1}u_{2}\rangle$ increase monotonically with $\lambda$ in weak
coupling regime. When $\lambda$ exceeds a critical value, $
-\langle n_{1}u_{1}\rangle$ and $-\langle n_{1}u_{2}\rangle$ show
different behaviors.  It is clearly shown that two electrons are
localized on one site in strong coupling region, resulting the
singlet-site bipolaron. It also demonstrates   that the e-e
interaction $U$ affects the magnitude of inter-site (on-site)
deformations $\langle n_{1}u_{2}\rangle$ ($-\langle
n_{1}u_{1}\rangle$) and the transition point where $-\langle
n_{1}u_{2}\rangle$ reaches maximum shifts to a larger $\lambda$ as
$U$ increases.

The nature of the bipolaron crossover can also be investigated by
the correlation function $\langle (n_1-n_2)(u_1-u_2)\rangle
/g_{1}$. Figs.\ref{correlation}(c) shows that the two-site
bipolaron regime of the coupling strength $\lambda$ is wider for
larger $U$ in the weak e-ph coupling regime. This crossover
behaviors are consistent with the above analysis on the  GS
entanglement and fidelity.

\section{summary and discussion}

In this present work we have solved exactly the two-site
Hubbard-Holstein model by the extended phonon coherent states
approach.  The GS energy is lower than previous results and the
crossover regime is revealed clearly by the electrons occupancy
probability. With the exact GS wave function, we study the crossover
properties from quantum information perspective based on the linear
entropy, the GS fidelity and its susceptibility. The single-site
bipolarons are maximally entangled and the bipolaron crossover
transition is more abrupt as e-e interaction $U$ increases. The
similar bipolaron crossover behaviors are also observed classically
in the static correlation function. The present study may provide
some insights into the more complicated Hubbard-Holstein systems in
a infinite chain.

\section{Acknowledgements}

This work was supported by National Natural Science Foundation of China,
PCSIRT (Grant No. IRT0754) in University in China, National Basic Research
Program of China (Grant No. 2009CB929104).

$^{\dag}$Corresponding author


\begin{thebibliography}{99}
\bibitem{ASAlexandrov}  A. S. Alexandrov, and N. F. Mott, polarons and
Bipolarons, World Scientific, Singapore (1995); A. Damascelli, Z. Hussain,
and Z. X. Shen, Rev. Mod. Phy. 75, 473 (2003).

\bibitem{AJMilis}  A. J. Millis, Nature (London) 392,147(1998); D. M.
Edwards, Adv. Phys. \textbf{51}, 1259 (2003).

\bibitem{THolstein}  T. Holstein, Ann. Phy. (NY) \textbf{8}, 325 (1959).

\bibitem{Fehske}  H. Fehske, G. Wellein, J. Loos, and A. R. Bishop, Phys.
Rev. B \textbf{77} , 085117 (2008).

\bibitem{Berciu}  M. Berciu, Phys. Rev. B \textbf{75}, 081101(R) (2007).

\bibitem{Scott}  S. Hill, and W. K. Wootters, Phys. Rev. Lett. \textbf{78},
26 (1997).

\bibitem{William}  W. K. Wootters, Phys. Rev. Lett. \textbf{80}, 10 (1998).

\bibitem{Vidal}  G. Vidal, J. I. Latorre, E.Rico, and A.Kitaev, Phys. Rev.
Lett. \textbf{90}, 227902 (2003).

\bibitem{Gu}  S. J. Gu, S. S. Deng, Y. Q. Li, and H. Q. Lin, Phys. Rev.
Lett. \textbf{93}, 086402 (2004).

\bibitem{chenqh} Q. H. Chen, Y. Y. Zhang, T. Liu, and K. L. Wang, Phys. Rev.
A (in press);  see also arXiv: 0809.4385.
\bibitem{Quan}  H. T. Quan, Z. Song, X. F. Liu, P. Zanardi, and C. P. Sun,
Phys. Rev. Lett. \textbf{96}, 140604 (2006).

\bibitem{Vezzani}  P. Buonsante, and A. Vezzani, Phys. Rev. Lett.
\textbf{98}, 110601 (2007).

\bibitem{Min}  M. F. Yang, arxiv:quant-ph/07074574

\bibitem{Shu}  S. Chen, L. Wang, Y. J. Hao, and Y. P. Wang, Phys. Rev. A.
\textbf{77}, 032111 (2008) .

\bibitem{Zanardi}  P. Zanardi, and N. Paunkovi\'{c}, Phys. Rev. E. \textbf{74%
}, 0331123 (2006).

\bibitem{Buon}  P. Buonsante, and A. Vezzani, Phys. Rev. Lett. \textbf{98},
110601 (2007).

\bibitem{Chen}  S. Chen, L. Wang, S. J. Gu, and Y. Wang, Phys. Rev. E.
\textbf{76}, 061108 (2007).

\bibitem{Dunning}  C. Dunning, J. Links, and H. Q. Zhou, Phys. Rev. Lett.
\textbf{94} , 227002 (2005).

\bibitem{Norman}  N. Oelkers, and J. Links, arxiv:cond-mat/0611510v2.

\bibitem{ANDas}  A. N. Das, and P. Choudhury, Phys. Rev. B \textbf{49}, 18
(1994).

\bibitem{Acquarone}  M. Acquaroneet al., Phys. Rev. B \textbf{58}, 7626
(1998).

\bibitem{JC}  J. Chatterjee, and A. N. Das, arXiv:cond-mat/0210607.

\bibitem{Ranninger}  J. Ranninger and U. Thibblin, Phys. Rev. B \textbf{45},
7730 (1992).

\bibitem{EVLde}  E. V. L. de Mello, and J. Ranninger, Phys. Rev. B \textbf{55%
}, 14872 (1997).

\bibitem{chen}  Q. H. Chen et al., Phys. Rev. B \textbf{53}, 11296(1996).

\bibitem{Han}  R. S. Han, Z. J. Lin, and K. L. Wang, Phys. Rev. B \textbf{65}%
, 174303(2002)

\bibitem{liu}  K. L. Wang, T. Liu, and M. Feng, Euro. Phys. J. B \textbf{54}%
, 283(2006).

\bibitem{Yang}  Y. Zhao, P. Zanardi, and G.H Chen, Phys. Rev. B. \textbf{70}%
, 195113 (2004).
\bibitem{Wang}  X. Wang, and B. Sanders, J. Phys. A. \textbf{38}, 67 (2005).

\bibitem{Cozzini}  M. Cozzini, P. Giorda, and P. Zanardi, Phys. Rev. B.
\textbf{75}, 014439 (2007).
\end{thebibliography}
\end{document}